\pdfoutput=1

\documentclass[10pt]{article} 
\usepackage{flushend}
\usepackage{setspace}
\usepackage{csquotes}
\usepackage{caption}
\usepackage{multirow}

\usepackage{subcaption}

\usepackage{endnotes}
\let\footnote\endnote
\def\footnotetext{\endnotetext[\number\numexpr\value{endnote}+1]}
\let\footnotemark\endnotemark

\usepackage{theapa}
\usepackage{amsmath,amssymb,amsthm}
\newtheorem*{definition}{Definition} 
\usepackage{enumitem}
\usepackage{multirow} 
\usepackage{titling}
\usepackage{url}
\usepackage[normalem]{ulem}
\usepackage{siunitx}
\pretitle{\LARGE\bfseries \sffamily}
\posttitle{}

  \renewenvironment{abstract}{%
       \bfseries \sffamily}{}
       \usepackage{titlesec}
\titleformat{\section}[block]{\vspace{-0.0em}\bfseries \large \sffamily}{\thesection.}{0em}{}[%
\nopagebreak\nopagebreak\vspace{0.0em}\nopagebreak ]

\titleformat{\subsection}[block]{\vspace{-0.0em}\slshape}{\thesection.}{0em}{}[ %
\nopagebreak\nopagebreak\vspace{0.0em}\nopagebreak ]

\usepackage{graphicx}
\usepackage{epstopdf}
\usepackage{color}
\usepackage{setspace}

\preauthor{\begin{flushleft}%
           \large \lineskip 0.5em%
           \begin{tabular}[t]{@{}l}}
\postauthor{\end{tabular}\end{flushleft}}
\renewcommand*\and{%
  \end{tabular}%
  \hskip 1em \relax
  \begin{tabular}[t]{l}}
\usepackage{caption}

\usepackage{fancyhdr} 
\renewcommand{\thispagestyle}[1]{}  
\begin{document} \pagestyle{fancy} \lfoot{\textit{To appear at JASIST.}}

\title{\noindent Measuring academic influence: \\
Not all citations are equal}

\author{%
\textbf{{\sffamily Xiaodan Zhu and Peter Turney}}\\
\emph{\small National Research Council Canada, Ottawa, ON K1A 0R6, Canada.} \\
\emph{\small 
Email: Xiaodan.Zhu@nrc-cnrc.gc.ca} 
\\
\textbf{{\sffamily Daniel Lemire}}\\
\emph{\small TELUQ, Universit\'e du Qu\'ebec, Montreal, QC H2S 3L5, Canada.} 
\\
\textbf{{\sffamily Andr\'e Vellino}}\\
\emph{\small School of Information Studies, University of Ottawa,
       Ottawa, ON K1N 6N5, Canada.} 
}
\date{}

\maketitle

\begin{abstract}

The importance of a research article is routinely measured by counting how many 
times it has been cited. However, treating all citations with equal weight ignores 
the wide variety of functions that citations perform. We want to automatically 
identify the subset of references in a bibliography that have a central academic 
influence on the citing paper. For this purpose, we examine the effectiveness of a 
variety of features for determining the academic influence of a citation.

By asking authors to identify the key references in their own work, we created a 
dataset in which citations were labeled according to their academic influence.
Using automatic feature selection with supervised machine learning, we found a 
model for predicting academic influence that achieves good performance on this
dataset using only four features.

The best features, among those we evaluated, were features based on the number of
times a reference is mentioned in the body of a citing paper. The performance of
these features inspired us to design an \emph{influence-primed \mbox{$h$-index}} 
(the hip-index). Unlike  the conventional \mbox{$h$-index}, it weights citations by 
how many times a reference is mentioned. According to our experiments, the hip-index
is a better indicator of researcher performance than the conventional \mbox{$h$-index}.

\end{abstract}
\section*{Introduction}

One of the functions of citation analysis is to determine the impact of an 
author's work on a research community. A first approximation for measuring 
this impact is to count the number of times an author is cited.  
Various other measures, such as the \mbox{$h$-index}~\cite{Hirsch:2005},  
the \mbox{$g$-index}~\cite{Egghe:2006} and the \mbox{$h_m$-index}~\cite{schreiber2008share}, 
refine this basic measure using functions based on the distribution of 
citations~\cite{bornmann2008there}.  

Yet other measures of author impact are based on methods for scoring articles 
with weights and thresholds that depend on the journals in which they were published and
the number of times the article was cited~\cite{marchant2009score}.
However, all these indexes and score-based rankings treat 
each citation as having equal significance.

It has long been recognized that not all citations are created equal and hence they
should not be counted equally. \citeA{readbeforeyoucite} estimated that authors read
only 20\% of the works they cite. This estimate was based on a detailed analysis of 
the frequency of replication of distinctive errors in citations, such as incorrect page
numbers or volume numbers. When an error in a citation is replicated many times, 
it seems likely that the citers have copied the citation without actually reading
the cited paper.

As an illustration, \citeA{saltonneverwrote} reported that a commonly cited paper 
by Gerard Salton does not actually exist. An incorrect citation was accidentally created by 
mixing the citations for two separate papers. This incorrect citation has since been 
cited by more than 300 papers. If the citers had tried to read the paper before citing 
it, they would have discovered that the paper does not exist. 

Like \citeA{Moravcsik:1975}, we are concerned about the side-effects of 
counting insignificant references:
\textquote[{\citeR[p.~91]{Moravcsik:1975}}]{%
A large fraction of the references are perfunctory. This raises serious doubts 
about the use of citations as a quality measure, since it is then quite possible 
for somebody or some group to chalk up high citation counts by simply writing 
barely publishable papers on fashionable subjects which will then be cited as 
perfunctory, `also ran' references.}. 
Indeed, based on an analysis of hundreds of references, \citeA{Moravcsik:1975} 
found that a third of the references were redundant and 40\% were perfunctory. 
In an independent study, \shortciteA{Teufel:2006} found that the majority (62.7\%) 
of the references could not be attributed a specific function whereas the fraction 
of references that provided an essential component for the citing paper 
(definition, tool, starting point)  was 18.9\%.

The aim of our work is to determine the most effective features for identifying  
references that have high \emph{academic influence} on the citing paper. 
An \emph{influential} reference is one that inspired a new idea, method, experiment, 
or research problem that is a core contribution of the citing paper.
We use the terms \emph{influence} and \emph{influential} to 
indicate the degree of academic influence of a single citation. In contrast, 
\citeA{pinski1976citation} used the term \emph{citation influence} to 
refer to  the academic influence of a journal.

Many attempts have been made to automatically identify which citations are most 
influential. Readers can often tell quickly whether a citation is shallow from 
the text itself, which has prompted several efforts to categorize citations by 
the linguistic context of their occurrence; that is, by the words near the citation 
in the body of the citing paper~\cite{Teufel:2006,Hanney:2005,Mercer:2004,Pham:2003}. 
\lfoot{}

In contrast to approaches based solely on linguistic context, our method uses 
machine learning to evaluate a number of citation features. We examine features 
based on linguistic context as well as other features, such as 

\begin{itemize}[itemsep=1pt,parsep=1pt,topsep=1pt,partopsep=1pt]

\item the location of the citation in the text, 
\item the semantic similarities between the titles of the cited papers and the content of 
the citing paper, 
\item the frequency with which the articles are cited in the 
literature, 
\item and the number of times a given reference is cited in the body of the paper.

\end{itemize}

We test the effectiveness of the features by applying machine learning to the 
problem of identifying the influential references. One of our most important 
contributions is to identify a set of four features that are particularly useful 
to determine influence. For example, the two best features are the number of 
times a reference appears and the similarity between the title of the cited 
paper and the core sections of the citing paper.

A secondary purpose of citation measures is to predict the future performance of 
authors, such as whether they will win a Nobel Prize~\cite{Garfield:1968,Gingras:2010}. 
The importance of researchers is reflected in the amount of influence they have on the 
research of their colleagues. Citation frequency is a measure of this influence, but a 
better measure would take into account \emph{how} a researcher is cited, rather than 
giving all citations equal weight. 

As a test of our method's ability to determine whether a cited paper substantially influenced 
the citing paper, we attempt to identify which researchers in computational linguistics 
are Fellows of the Association for Computational Linguistics (ACL), based solely on their
publication records and citations. For each ACL Fellow, we compare their conventional 
(unweighted) \mbox{$h$-index} with an \mbox{$h$-index} computed from citations weighted 
by our measure of academic influence. We get a better average precision measure with 
weighted citations.

\section*{Defining academic influence}

What does it mean to say that one reference had more academic influence on
a given citing paper than another reference? If our aim is to distinguish references
according to their degrees of academic influence, then we must be precise about
the meaning of academic influence. As researchers, we know that some papers have
influenced the course of our research more than others, but how can we pin down
this intuition?

A paper written by an evolutionary biologist is likely to have been influenced
by Darwin's \emph{On the Origin of Species}, but we are more interested in the
\emph{proximate} influences on the paper. A good research paper contributes a new
idea to the literature. What prior work was the proximate cause, the impetus
for that new idea?

We believe that this question is best answered by the authors of the citing paper, 
because they are in the best position to decide which of their references should be 
labeled \emph{influential} and which should be labeled \emph{non-influential}. 
It could be said that we avoid the problem of precisely defining influence; instead 
we give a kind of \emph{operational definition}: A cited paper is influential for a given citing
paper if the authors of the citing paper say that it is influential.

We acknowledge that authors may be wrong about whether a paper was influential.
Two types of error are possible: Authors may say a reference is \emph{influential}
when it is actually \emph{non-influential} or authors may say a reference
is \emph{non-influential} when it is actually \emph{influential}. Both types
seem plausible to us. In the first case, the authors might feel obliged to
say that a paper is \emph{influential}, because the paper is
very popular, very respected, or very well written. In the second case,
a paper might have greatly influenced the authors at a subconscious level, but
they might mistakenly say it is \emph{non-influential}, or they might not want to
admit that there was any influence, due to professional jealousy. 
Nevertheless, although the authors might be wrong, we know of no better, more reliable 
way of determining which references were influential. Therefore we base our experiments 
on author-labeled data.

\citeA{dietz2007unsupervised} also rely on author-labeled data. They
collected a data set of twenty-two papers labeled by their authors. Each reference was labeled 
on a Likert scale and they experimented with unsupervised prediction of citation influence. 
Unlike us, their purpose was to model  topical inheritance via citations. 

\section*{Motivation}

Suppose that we have a model for predicting the label (\emph{influential} or \emph{non-influential}) 
of a paper--reference pair, consisting of a given citing paper and a given citation 
within that citing paper. We label pairs rather than references alone, because a reference 
that  is influential for one citing paper is not necessarily influential for another
citing paper. Such a model would have many potential applications.
Wherever citation counts play an important role, the model could be applied
to filter or weight the citations. Some potential applications follow.

Summarizing: Given a paper with a long list of
references, the model could identify the most influential references
and list them. For those who are familiar with the field of the given
paper, this list would rapidly convey the topic and nature of the paper.
For those who are new to the field, this list would suggest further
reading material. Citations have generally proven useful for  
summarization~\cite{Qazvinian:2008:SPS:1599081.1599168,Qazvinian:2010:INC:1858681.1858738,Abu-Jbara:2011:CCS:2002472.2002536,nanba1999towards,Taheriyan:2011:SCR:2023568.2023579,Kaplan:2009:AEC:1699750.1699764}.

Improved measures of an author's impact: Indexes such as the \mbox{$h$-index}~\cite{Hirsch:2005},  
\mbox{$g$-index}~\cite{Egghe:2006}, and \mbox{$h_m$-index}~\cite{schreiber2008share}
could be made less sensitive to noise by filtering citation counts with a model of influence. 
Beyond reducing sensitivity to noise, a model of influence could also put more weight on original 
contributions. It is known that survey papers and methodology papers tend to be more highly cited
than research contributions in general~\cite{ioannidis2006concentration}. 
\citeA{vanclay2013factors} went as far as to recommend focusing on reviews:
\textquote[{\citeR[p.~270]{vanclay2013factors}}]{%
Perhaps the best single thing an aspiring author can do to attract citations is to 
participate in a rigorous review rather than writing a conventional research article.}.
However, survey and methodology papers seem less likely to us to be labeled 
as influential by authors. Filtering citation counts by a model of influence might decrease the 
impact of survey and methodology papers, putting more weight on innovative research. 

Improved journal impact factors: As with measures of
author impact, measures of journal impact~\cite{bollen2009principal} 
could be made less sensitive to noise by filtering citation
counts with a model of influence. 

Improved measures of research organization impact:
Citations counts are also used to evaluate research organizations.
As with journal impact and author impact, performance
measures that are based on citation counts may benefit
from filtering by a model of influence.

Meme tracking: Historians of science are interested in tracking the spread of ideas 
(memes)~\cite{Haque:2011:PSC:1998076.1998081,Leskovec:2009:MDN:1557019.1557077}. 
Citations are a noisy way to track how ideas spread, because a reference
may be cited for many reasons other than being the source
of an influential idea~\cite{bornmann2008citation}. Filtering by a model of influence
may result in a better analysis of the spread of an idea.

Research network analysis: Scientists belong to networks
of people who collaborate with each other or influence
each other's work. Filtering citations with a model of influence
may make it easier to identify these networks automatically.

Improved hyperlink analysis: In many ways, hypertext links in
web pages are analogous to citation links in research papers.
A good model of citation influence could suggest a model
of hypertext link importance. This could improve measures
of the importance of web pages, such as PageRank~\cite{Qi:2007:MSD:1244408.1244418}. 

Improved recommender systems: Researchers often need help 
identifying relevant work that they should read. Filtering out less 
relevant citations might help paper recommender 
systems~\cite{MEET:MEET14504701330,springerlink:10.1007/978-3-642-23535-1_35}.

\section*{Related work}

The idea that the mere counting of citations is dubious is not new~\cite{chubin1975content}: 
The field of \emph{citation context analysis} \cite<a phrase coined by>{Small:1982} has a long 
history dating back to the early days of citation indexing. There is a wide variety of reasons 
for a researcher to cite a source and many ways of categorizing them.
For instance, \citeA{Garfield:1965} identified fifteen such reasons, including giving credit 
for related work, correcting a work, and criticizing previous work. 

For articles in the field of high energy physics, \citeA{Moravcsik:1975} distinguished four 
major classes of polar opposite pairs, conceptual--operational, organic--perfunctory, 
evolutionary--juxtapositional, and confirmative--negational. 
They found that the fraction of negational references, i.e., citations indicating that 
the cited source is wrong, is not negligible (14\%).

\citeA{Giles:1998:CAC:276675.276685} presented one of the first  automatic citation indexing 
systems (CiteSeer). It could parse citations and use them to compute similarities between documents.

\citeA{garzone2000towards} might have implemented the first automated classification systems 
for citations. They used over 200~manually selected rules to classify citations in one of 
35~categories.

Machine learning methods for automatic classification can be applied to the text of a 
citing document. \citeA{Teufel:2006} distinguish categories of citations that can be identified 
via linguistic cues in the text. They are able to classify citations into one of four categories 
(weak, positive, contrast, neutral) with an average \mbox{F-measure} of 68\%. For a classification in three
categories (weakness, positive, neutral), they get an average \mbox{F-measure} of 71\%. Their classifier 
relies on 892~manually selected cue phrases, such as whether the citation is a self-citation, the location 
of the citation in the text, and manually acquired verb clusters.

\citeA{agarwal2010automatically} annotated a corpus of 43~open-access full-text biomedical 
articles. They built classifiers using Support Vector Machine (SVM) and Multinomial Na\"ive Bayes 
(MNB) models using the  open-source Java library Weka. They report an average \mbox{F-measure} of 76.5\%. 
They used unigrams (individual words) and bigrams (two consecutive words) as features. They ranked 
their features using mutual information. They found the SVM models were generally superior to
the MNB models.

Our own methodology differs from \citeA{Teufel:2006} and \citeA{agarwal2010automatically} 
in at least one significant way: We asked the authors of the citing papers themselves to identify 
the influential references whereas they used independent annotations. We believe that it is difficult for an
independent annotator to classify citations. This concern was raised by \citeA{gilbert1977referencing}: 
\textquote[{\citeR[p.~120]{gilbert1977referencing}}]{%
Since the intentions of the author are not normally available to the content analyst, there seems 
to be no way of conclusively resolving problems of classifications (\ldots) The difficulties are 
more compounded when the analyst has only a superficial knowledge of the contexts in which the 
papers he examines were written and read.}.  
Nevertheless, \citeA{Teufel:2006} report moderately good inter-annotator agreement.

To address concerns about the consistency of the \mbox{$h$-index}, ranking and scoring have been proposed 
as alternative measures \cite{waltman2012inconsistency}. Like the \mbox{$h$-index}, these
measures are based on citation counts, and they too could benefit from 
filtering or weighting citations with a model of influence. 

There are other good reasons, beside assessing researchers, to make distinctions between 
different types of citations. The need also arises from the desire by publishers to provide 
scholarly research with semantic annotations. Thus CiTO, the Citation Typing 
Ontology~\cite{Peroni:2012,shotton2010cito}, provides a rich machine-readable RDF metadata 
ontology for the characterization of bibliographic citations.

From among the almost ninety semantic relations for citations identified in CiTO (for example, 
\emph{agrees with}, \emph{obtains background from}, \emph{supports}, and \emph{uses conclusions from}), 
it is natural to generalize at least two broad categories  of citations, ones that acknowledge a 
fundamental intellectual legacy (such as \emph{critiques}, \emph{extends}, and \emph{disputes}) 
and ones that are incidental (such as \emph{cites for information}, \emph{obtains support from}, 
and \emph{cites as related}).

Several authors have proposed weighting citations based on factors such as the prestige of 
the citing journal~\cite{ding2011applying,yan2010weighted,ding2011popular,gonzalez2010new}. 
Others have proposed weighting citations by the  mean number of references of the citing 
journal~\cite{zitt2008modifying}. \citeA{balaban2012positive} proposed giving more weights to 
citations from more prestigious authors. He also argues that the citation of a paper published in 
a less prestigious venue should be considered more significant:
\textquote[{\citeR[p.~244]{balaban2012positive}}]{%
\dots  if a paper is cited despite its handicap of having appeared in a low-IF [Impact Factor] 
journal, then this means that this paper has a high intrinsic value. Therefore a citation's 
value should be inversely correlated with the IF of the journal in which the cited paper was 
published.}.

\citeA{BIES:BIES201100067} propose to use frequency to assess the importance of a citation. 
That is, if a reference was cited 10 times in the citing paper it gets a weight of 10. They 
show that by weighting citations by the in-paper frequency, review articles lose part of 
their advantage over original contributions when counting citations. They state that 
greater credit is reverted to the discoverers. They also show that closely related
references are cited more often in the body of a citing paper than less related references
(on average, 3.35~times versus 1.88~times). They define \emph{closely related references} 
as papers having at least ten references in common with the given paper.

\citeA{marco2006using} studied hedging as a means to classify citations.
\citeA{springerlink:10.1007/11731139_32} used finite-state machines for classification of citations.

\section*{Features for supervised learning}
\label{sec:features}

We are concerned with a binary classification problem: Given a
research paper, classify its references as either \emph{influential}
or \emph{non-influential}. 

The task is to create a model that takes a pair, consisting of a given author's paper (the citer)
and a reference in the given paper (the cited paper), as input, and generate
a label (influential or non-influential) as output. Our goal is to create
a model that can predict the labels assigned by the authors in our gold-standard
dataset. Our approach to this task is to use supervised machine learning.

For supervised machine learning, we must generate feature vectors that represent
a variety of properties of each paper--reference pair. Given a training set of 
manually labeled paper--reference feature vectors, a learning algorithm can create 
a model for predicting the label of a paper--reference pair in the testing set.

We use a standard supervised machine learning algorithm in our experiments
(a support vector machine). The main contribution of this paper is that
we evaluate a wide range of different features for representing the 
paper--reference pairs. Finding good features is the key to successful prediction.

In our experiments, we consider five general classes of features:

\begin{enumerate}[itemsep=1pt,parsep=1pt,topsep=1pt,partopsep=1pt]

\item Count-based features
\item Similarity-based features
\item Context-based features
\item Position-based features
\item Miscellaneous features

\end{enumerate}

\noindent Not all of these features are useful. However, many of these features are 
intuitively attractive, and we can only find out if (and the extent to which) they 
are useful through experiments. We describe these features in the following subsections.

\subsection*{Count-based features}

The count-based features are based on the intuition that a reference
that is frequently mentioned in the body of a citing paper is more
likely to be influential than a reference that is only mentioned once.
We created five different count-based features:

\begin{enumerate}[itemsep=1pt,parsep=1pt,topsep=1pt,partopsep=1pt]

\item Count-based features

\begin{enumerate}[itemsep=1pt,parsep=1pt,topsep=1pt,partopsep=1pt,label*=\arabic*.]

\item countsInPaper\_whole
\item countsInPaper\_secNum
\item countsInPaper\_related
\item countsInPaper\_intro
\item countsInPaper\_core

\end{enumerate}
\end{enumerate}

We count the occurrences of each reference in the
entire citing paper ($counts\allowbreak{}InPaper\allowbreak{}\_whole$), in the introduction
($countsInPaper\_intro$), in the related work ($countsInPaper\_related$),
and in the core sections ($countsInPaper\_core$), where \emph{core
sections} include all the other sections, excluding those already mentioned 
and excluding the acknowledgment, conclusion, and future-work
sections. 

We added a feature ($countsInPaper\_secNum$) to indicate the number of 
different sections in which a reference appears. This feature is based on 
the intuition that a reference that appears in several different
sections is more significant than a reference that appears in only
one section (even if it may have a high frequency within that one section).

\subsection*{Similarity-based features}

It seems natural to suppose that the influence of a cited paper
on a given citing paper (the citer) is proportional to the overlap
in the semantic content of the cited paper and the citer. That is,
if there is a high degree of semantic similarity between the text
of the citer and the text of the cited paper, then it seems likely that 
the cited paper had a significant influence on the citer.
Accordingly, we explored a variety of features
that attempt to capture semantic similarity.

We assume that the text of the citer is given, but we do not assume
that we have access to the text of the cited paper. The cited paper
might not be readily available online, due to subscription charges
or the age of the paper. A benefit of our approach is 
that all these features can be implemented efficiently: All features
can often be computed from a single document. Even when the
titles of references are not included by the journal, it may
still be easier to locate the missing titles than the full text.

Since (by choice) we do not have  access to the full text of the cited paper, 
we use the title of the cited paper as a surrogate for the full text.
The first five similarity-based features compare the title of the
cited paper to various parts of the citing paper:

\begin{enumerate}[itemsep=1pt,parsep=1pt,topsep=1pt,partopsep=1pt,resume]

\item Similarity-based features

\begin{enumerate}[itemsep=1pt,parsep=1pt,topsep=1pt,partopsep=1pt,label*=\arabic*.]

\item sim\_titleTitle
\item sim\_titleCore
\item sim\_titleIntro
\item sim\_titleConcl
\item sim\_titleAbstr

\end{enumerate}
\end{enumerate}

We calculate the similarities between the title of a reference and the
title ($sim\_titleTitle$), the abstract ($sim\_titleAbstr$), the
introduction ($sim\_titleIntro$), the conclusion ($sim\_titleConcl$),
and the core sections ($sim\_titleCore$) of the citing paper. These
features should be able to capture the semantic similarity between the
citer and a reference; in most cases, a title, abstract, and
conclusion section are good summaries of the given citing
paper. \textit{Core sections} here refer to the same sections as in
the count-based features.

More specifically, we calculated cosine similarity scores. A piece of
text (e.g., a title or abstract) is first represented as a vector in
the word space, where each dimension is a word (a word type; not a word token). 
The values in the vector are the word frequencies of each word appearing in this
piece of text. \citeS{Porter:1980} stemming algorithm was used to stem
the text (remove suffixes) and we kept stop words (function words), since removing 
them did not improve the performance of our models during their development. Readers can
refer to \citeA{Turney:2010} for further discussions of vector space
models of semantic similarity.

In a given citing paper (the citer), when a reference is mentioned
in the body of the citer, the text that appears near the mention
is called the \emph{citation context}. Like the title of the cited paper,
the citation context provides information about the cited paper; hence
we can use the citation context as a surrogate for the full text of the
cited paper, in the same way that we used the title as a surrogate.
The next four similarity-based features compare the citation context
to various parts of the citing paper:

\begin{enumerate}[itemsep=1pt,parsep=1pt,topsep=1pt,partopsep=1pt,resume,start=2]

\item Similarity-based features

\begin{enumerate}[itemsep=1pt,parsep=1pt,topsep=1pt,partopsep=1pt,resume,start=6,label*=\arabic*.]

\item sim\_contextTitle
\item sim\_contextIntro
\item sim\_contextConcl
\item sim\_contextAbstr

\end{enumerate}
\end{enumerate}

For each reference, we calculate the similarities between the citation contexts 
and the title ($sim\_contextTitle$), abstract ($sim\_contextAbstr$), introduction
($sim\_contextIntro$), and conclusion ($sim\_contextConcl$) of the citing
paper. When a reference appears multiple times in the citer, we take
the average of the similarities over all its contexts. 

As with title similarity (features 2.1 to 2.5), we use cosine similarity, after 
the text was preprocessed with \citeS{Porter:1980} stemmer. During development, we
experimented with different window sizes, ranging from two words
around a citation to several sentences around it. We found that using
the entire sentence in which the citation appears gave the best results.
In contrast, \citeA{Ritchie:2008:CCC:1458082.1458113} found that contexts larger than one 
sentence were better for indexing purposes: further work might
be needed to identify the optimal window.

\subsection*{Context-based features}

The citation context of a reference could indicate the academic
influence of the reference in other ways, beyond its value as
as a surrogate for the full text of the cited paper (as in
the above features 2.6 to 2.9). For example, if a citation $X$ appears
in the context ``the work of $X$ inspired us'', then $X$ 
seems likely to be influential for the given citing paper.

For these features, we define the citation context to be a window of 
ten words around a citation (five words on each side). If a reference 
appears multiple times in the citing paper, we calculate its average score.

The first three context-based features are based on the relation
between the citation and the citation context:

\begin{enumerate}[itemsep=1pt,parsep=1pt,topsep=1pt,partopsep=1pt,resume]

\item Context-based features

\begin{enumerate}[itemsep=1pt,parsep=1pt,topsep=1pt,partopsep=1pt,label*=\arabic*.]

\item contextMeta\_authorMentioned
\item contextMeta\_appearAlone
\item contextMeta\_appearFirst

\end{enumerate}
\end{enumerate}

The first feature ($contextMeta\_authorMentioned$) indicates whether the authors 
of a reference are explicitly mentioned in the citation context; for example, 
``the work of Smith et al.\ [4]'' mentions the authors (Smith et al.) in the citation 
context of the reference ([4]). The second feature ($contextMeta\_appearAlone$) indicates 
whether a citation is mentioned by itself (e.g., ``[4]'') or together with other citations 
(e.g., ``[3,4,5]''). When a citation is mentioned with other citations, the third
feature indicates whether it is mentioned first (e.g., ``[4]'' is first in ``[4,5,6]'').

These three features may be biased by the different citation format requirements of various
journals, but we leave it to the supervised learning system to decide whether the
features are useful. A feature may be useful for prediction even when it has some bias.
(However, we will see later that these features were not particularly effective in our
experiments.)

The next twelve context-based features are based on the meaning of 
the words in the citation context:

\begin{enumerate}[itemsep=1pt,parsep=1pt,topsep=1pt,partopsep=1pt,resume,start=3]

\item Context-based features 

\begin{enumerate}[itemsep=1pt,parsep=1pt,topsep=1pt,partopsep=1pt,resume,start=4,label*=\arabic*.]

\item contextLex\_relevant
\item contextLex\_recent
\item contextLex\_extreme
\item contextLex\_comparative
\item contextLexOsg\_wnPotency
\item contextLexOsg\_wnEvaluative
\item contextLexOsg\_wnActivity
\item contextLexOsg\_giPotency
\item contextLexOsg\_giEvaluative
\item contextLexOsg\_giActivity
\item contextLexEmo\_emo
\item contextLexEmo\_polarity

\end{enumerate}
\end{enumerate}

We manually created four relatively short lists of words that we designed to
detect whether the citation context suggests that the cited paper
is especially relevant to the citer ($contextLex\_relevant$),
whether the citation context signals that the cited paper is new
($contextLex\_recent$), whether the citation context implies
that the cited paper is extreme in some way ($contextLex\_extreme$), and
whether the citation context makes some kind of comparison with
the cited paper ($contextLex\_comparative$). The names of
these features convey the kinds of words in the short lists. In  
Table~\ref{table:shortlists}, we give the full lists for $contextLex\_relevant$  
and $contextLex\_recent$ and a few terms for the other two features (as they 
both contain over 100~words).

\begin{table}
\caption{Manually created  lists of words to classify the citation context 
\label{table:shortlists}}
\centering
\begin{tabular}{lp{0.6\textwidth}}\hline
$contextLex\_relevant$ & \footnotesize relevant, relevantly, related, relatedly, 
similar, similarly, likewise, pertinent, applicable, appropriate, useful, pivotal, 
influential, influenced, comparable, original, originally, innovative, suggested, 
interesting, inspiring, inspired \\
$contextLex\_recent$ & \footnotesize recent, recently, up-to-date, latest, 
later, late, latest, subsequent, subsequently, previous, previously, initial, 
initially, continuing, continued, sudden, current, currently, future, unexpected, 
upcoming, expected, ongoing, imminent, anticipated, unprecedented, proposed, 
startling, preliminary, ensuing, repeated, reported, new, old, early, earlier, 
earliest, existing, further, update, renewed, revised, improved, extended \\
$contextLex\_extreme$ & \footnotesize greatly, intensely, acutely, almighty,
awfully, drastically, exceedingly, exceptionally, excessively, \ldots \\
$contextLex\_comparative$ & \footnotesize easy, easier, easiest, strong, 
stronger, strongest, vague, vaguer, vaguest, weak, weaker, weakest, \ldots \\\hline
\end{tabular}
\end{table}

We also created features based on \citeS{Osgood:1957} semantic differential categories 
($contextLexOsg$). \citeA{Osgood:1957} discovered that three main factors accounted 
for most of the variation in the connotative meaning of adjectives. 
The three factors were \emph{evaluative} (good--bad), \emph{potency} (strong--weak), 
and \emph{activity} (active--passive). 

The General Inquirer lexicon \cite{stone1966general}
represents these three factors using six labels, \emph{Positiv} and \emph{Negativ}
for the two ends of the \emph{evaluative} continuum, \emph{Strong} and \emph{Weak}
for the two ends of the \emph{potency} continuum, and \emph{Active} and \emph{Passive}
for the two ends of the \emph{activity} continuum.\footnote{See
\url{http://www.wjh.harvard.edu/~inquirer/} to obtain a copy of the 
General Inquirer lexicon.}

The feature $contextLexOsg\_giEvaluative$ is the number of words in the citation 
context that are labeled \emph{Positiv} in the General Inquirer lexicon, 
$context\allowbreak{}LexOsg\allowbreak{}\_giPotency$ is the number of words labeled \emph{Strong}, and 
$context\allowbreak{}LexOsg\allowbreak{}\_giActivity$ is the number of words labeled \emph{Active}. 
The intuition behind these features is that a citation is more likely to be influential 
if positive, strong, active words occur in the citation context.

The General Inquirer lexicon has labels for 11,788~words. Using the algorithm of 
\citeA{Turney:2003}, we automatically extended the labels to cover 114,271 words. The 
additional words are from the WordNet lexicon. The WordNet features ($contextLexOsg\_wn$) 
are similar to the corresponding General Inquirer features ($contextLexOsg\_gi$), except 
they include these additional words.\footnote{See \url{http://wordnet.princeton.edu/} to download 
the WordNet lexicon.}

Since the citation context is a window of ten words around the citation, the values
of these features range from zero to ten. If a reference is cited multiple times in the 
body of the citing paper, we calculate its average value. For increased precision,
we only considered the words in the citation context that have an adjective or adverb 
sense in WordNet. 

We used an emotion lexicon \cite{Mohammad:2010} to check whether the citation
context includes words that convey sentiment ($context\allowbreak{}LexEmo\allowbreak{}\_polarity$) or 
emotion ($context\allowbreak{}LexEmo\allowbreak{}\_emo$). The lexicon contains human annotation of
emotion associations for about 14,200~words. The annotations in the lexicon
indicate whether a word is positive or negative (known as \emph{sentiment},
\emph{polarity}, or \emph{semantic orientation}), and whether it is associated 
with eight basic emotions (joy, sadness, anger, fear, surprise, anticipation, 
trust, and disgust). 

The feature $contextLexEmo\_polarity$ is the number of
words in the citation context that are labeled either positive or negative.
The feature $context\allowbreak{}LexEmo\allowbreak{}\_emo$ is the number of words that are labeled with
any of the eight basic emotions. The idea behind these features is that
any kind of sentiment or emotion in the words in the citation context might
indicate that the citation is influential, even if the sentiment or emotion
is negative.

As with the other $contextLex$ features, the values of $contextLexEmo$ range
from zero to ten. Multiple occurrences of a citation are averaged.

\subsection*{Position-based features}

The location of a citation in the body of a citing paper might
be predictive of whether the cited paper was influential.
Intuitively, the earlier the citation appears in the text, the
more important it seems to us. The first two types of position-based features are 
based on the location of a citation in a sentence:

\begin{enumerate}[itemsep=1pt,parsep=1pt,topsep=1pt,partopsep=1pt,resume]

\item Position-based features

\begin{enumerate}[itemsep=1pt,parsep=1pt,topsep=1pt,partopsep=1pt,label*=\arabic*.]

\item posInSent\_begin
\item posInSent\_end

\end{enumerate}
\end{enumerate}

These are binary features indicating whether a citation appears at the beginning
($posInSent\_begin$) or the end ($posInSent\_end$) of the sentence. If
a reference appears more than once in the citing paper, we calculated
the percentages; for example, if a reference is cited three times in the paper
and two of the three appear at the beginning of the sentences, the
$posInSent\_begin$ feature takes the value of 0.667. 

The next four position-based features are based on the location
of a citation in the entire citing paper:

\begin{enumerate}[itemsep=1pt,parsep=1pt,topsep=1pt,partopsep=1pt,resume,start=4]

\item Position-based features

\begin{enumerate}[itemsep=1pt,parsep=1pt,topsep=1pt,partopsep=1pt,resume,start=3,label*=\arabic*.]

\item posInPaper\_stdVar
\item posInPaper\_mean
\item posInPaper\_last
\item posInPaper\_first

\end{enumerate}
\end{enumerate}

We measured the positions of the sentences that cite a given reference, 
including the mean ($posInPaper\_mean$), standard variance ($posInPaper\_stdVar$),
first ($posInPaper\_first$), and last position ($posInPaper\_last$) of
these sentences. These features are normalized against the total
length (the total number of sentences) of the citing paper; thus
the position ranges from 0 (the beginning of the citing paper) to 1
(the end of the citing paper).

More sophisticated location-based features are possible but not considered. For example,
references appearing  in a methodology section might be more influential
than those appearing solely in the related work section. 

\subsection*{Miscellaneous features}

The next three features do not fit into the previous four classes
and they have little in common with each other. We arbitrarily put
them together as \emph{miscellaneous} features:

\begin{enumerate}[itemsep=1pt,parsep=1pt,topsep=1pt,partopsep=1pt,resume]

\item Miscellaneous features

\begin{enumerate}[itemsep=1pt,parsep=1pt,topsep=1pt,partopsep=1pt,label*=\arabic*.]

\item aux\_citeCount
\item aux\_selfCite
\item aux\_yearDiff

\end{enumerate}
\end{enumerate}

The citation count a paper has received (in the general literature; not the
number of occurrences within a specific paper) is widely used as a metric
for estimating the academic contribution of a paper, which in turn is an
essential building block in calculating other metrics (e.g., \mbox{$h$-index})
for evaluating the academic contribution of a researcher, organization, or
journal. We are interested in understanding its usefulness in deciding academic
influence (in a specific paper). That is, when cited in a given paper, is a more 
highly cited paper more likely to have academic influence on the citer? To explore this
question, we collected the raw citation counts of each reference in Google Scholar
($aux\_citeCount$).

In accordance with convention, self-citation refers to the phenomenon where a citer
and a reference share at least one common author. We are interested in
knowing whether a self-citation would have a positive or negative
correlation with academic influence. To study this, we manually
annotated self-citation among the references and used it as a binary
feature ($aux\_selfCite$).

Are older papers, if cited, more likely to be academically influential? We incorporated 
the publication year of a reference as a feature ($aux\_yearDiff$). We calculated 
the difference in publication dates between a reference and the citer by subtracting 
the former from the latter, which resulted in a non-negative integer feature.

\subsection*{Contextual normalization}

Many of the above features are sensitive to the length of the citing paper.
For example, the number of occurrences of each reference in the
entire citing paper ($countsInPaper\_whole$) tends to range over larger
values in a long paper than in a short paper. For predicting whether a reference
is influential, it is useful to normalize the raw feature values for
a given paper--reference pair by considering the range of values in the
given citing paper. This is a form of \emph{contextual normalization}~\cite{Turney:1993,Turney:1996},
where the citing paper is the context of a feature.

We normalize all our features so that their values are in the range $[0,1]$.
This kind of normalization is standard practice in data mining, as it improves
the accuracy of most supervised learning algorithms \cite{witten2011data}. For example, 
consider the feature $countsInPaper\_whole$, where we count how many times
a given reference is cited in the whole text. Suppose we find that, in a given 
citing paper, one reference is cited ten times, but all other references are cited only
once. We would then give a score of 1 to the most cited reference, and a score of $1/10$ 
to the other references. That is, the most often cited references in any given citing paper 
always get a score of 1.

We formalize the normalization as follows. Our feature set contains both binary and 
real-valued features that take non-negative values. Binary features do not require 
normalization: Their values are 0 and 1.\footnote{This normalization 
leaves binary values unchanged, so it makes no difference whether it is applied to them.}
Other features are normalized to $[0,1]$. Let $\langle p_i, r_{ij} \rangle$ be a paper--reference pair, 
where $p_i$ is the \mbox{$i$-th} citing paper and $r_{ij}$ is the \mbox{$j$-th} reference in $p_i$.
Let $f_k$ be the \mbox{$k$-th} feature in our feature set and let $v(p_i,r_{ij},f_k)$
be the value of the feature $f_k$ in the paper--reference pair $\langle p_i, r_{ij} \rangle$.
Suppose that $p_i$ contains $n$ distinct references, 
$\langle p_i, r_{i1} \rangle, \dots, \langle p_i, r_{in} \rangle$,
resulting in $n$ values for $f_k$, $v(p_i,r_{i1},f_k), \dots, v(p_i,r_{in},f_k)$.
Let $\max (p_i,r_{i*},f_k)$ be the maximum of the $n$ values,
$v(p_i,r_{i1},f_k), \dots, v(p_i,r_{in},f_k)$.
We normalize each $v(p_i,r_{ij},f_k)$ to range from zero to one, using the
formula $v(p_i,r_{ij},f_k) / \allowbreak{}\max (p_i,r_{i*},f_k)$.
If $\max (p_i,r_{i*},f_k)$ is zero, then we normalize $v(p_i,r_{ij},f_k)$
to zero.

\section*{Experiments with features}

Using a labeled dataset, we first identify the features that are most
correlated with academic influence. We then combine some of these features to 
achieve a good classification score. 

\subsection*{Gold-standard dataset}

We believe that the authors of a paper are in the best position
to determine whether a given reference had a strong influence
on their research. In a blog posting, we invited authors to help us create a \emph{gold-standard
dataset} of labeled references.\footnote{See \url{http://tinyurl.com/counting-citations}.}
The authors were directed to fill in an online form.\footnote{See 
\url{http://tinyurl.com/influential-references}.} The instructions on the form were as follows:

\begin{quote}
We believe that most papers are based on 1, 2, 3 or 4~essential references. 
By an essential reference, we mean a reference that was highly influential or 
inspirational for the core ideas in your paper; that is, a reference that inspired 
or strongly influenced your new algorithm, your experimental design, or your choice 
of a research problem. Other references merely support the work. 

We believe that authors are the best experts to assess which references are essential. 
We are interested in automatically finding these references. To know how well we are 
doing, we need your help: please give us the title of a few of your papers and list 
for each paper the references that you feel are most essential, those without which 
the work would not have been possible.
\end{quote}

Forty different researchers filled out our online form (see Table~\ref{table:vol}). 
About half of them are from the USA and Canada. Three quarters of them are in computer science.
\begin{table}
\caption{\label{table:vol}Volunteer researchers}
\centering
  \subcaptionbox{Geographical provenance}{%
  \hspace*{1cm}
\begin{tabular}{lr}
\hline
Country & Number \\
\hline
Belgium & 1\\
Brazil& 1\\
Canada & 9 \\
Denmark & 1\\
France & 4 \\
Germany &2\\
Greece & 1 \\
Norway &1\\
Poland & 1 \\
Portugal & 1\\
Singapore & 1\\
Slovenia & 1\\
Spain& 1\\
UK & 3 \\
USA& 12 \\
\hline
\end{tabular}	
\hspace*{1cm}
}\subcaptionbox{Research domain}{%
\hspace*{1cm}\begin{tabular}{lr}
\hline
Discipline & Number \\
\hline
Biophysics & 1 \\
Chemistry & 1 \\
Computer Science & 30\\
Ecology &  1\\
Genetics &  2\\
Geophysics & 1\\  
Mathematics& 1\\
Physics& 1\\
Signal Processing& 1\\  
Translation& 1\\  
\hline
\end{tabular}	
\hspace*{1cm}
}
\end{table}

This gold-standard dataset provides us with a benchmark for supervised
machine learning.\footnote{The dataset is freely available online at 
\url{http://lemire.me/citationdata/}.} The authors gave us the titles of their 
papers and they indicated which references in each paper were influential for them.
From the titles, we obtained PDF copies of their papers and converted them
to plain text. We then extracted the references from the text and labeled
them as influential or non-influential. 

In total, the authors contributed 100~of their papers. OpenNLP was
used to detect sentence boundaries and conduct
tokenization.\footnote{See \url{http://opennlp.apache.org/} to
download OpenNLP.} We then used ParsCit to parse the
papers \cite{Councill:2008}. ParsCit is an open-source package for
parsing references and document structure in scientific papers. We
first ran the papers through ParsCit and then used a few hand-coded
regular expressions to capture citation occurrences in paper bodies
that were not detected by ParsCit.

The contents of the papers were then further annotated. First, the
section names were standardized to twelve predefined labels:
\emph{title}, \emph{author}, \emph{abstract}, \emph{introduction}, \emph{related}, 
\emph{main}, \emph{conclusion}, \emph{future}, \emph{acknowledgment}, 
\emph{reference}, \emph{appendix}, and \emph{date} (the year of publication
of the given paper). The default label for a section was \emph{main} (the core 
or main body of the paper). For example, \emph{previous work} and \emph{related work} 
would both be standardized to \emph{related}.

Second, the bibliographic items were manually corrected and meta-data about 
them (e.g., the Google citation counts) was included. The citations of these 
items in the main body of the paper were also manually corrected. For example, 
if references were cited as ``[7-10]'', we modified the citation to 
\mbox{``[7, 8, 9,10]''}, so as to explicitly include references [8] and [9]. 

\citeS{Porter:1980} stemmer (mentioned in the preceding section) was only applied
as a preprocessing step when generating feature vectors; the stemmer was
not applied during the corpus annotation step described here. 

The basic units in our study are paper--reference pairs, not papers.
The 100~papers yield 3143~paper--reference pairs (that is, 3143~data points;
3143~feature vectors). In the main bodies of the 100~papers, there
are 5394 occurrences of the references, so each paper contains an average of 
around 31 references (in the bibliographies) and 54~citations (in the main
text). The dataset contains 322 (10.3\% of 3143) \emph{influential} references
(strictly speaking, 322 influential paper--reference pairs). That is, their 
authors identified an average of 3.2~influential references per research paper.

\subsection*{Correlation between labels and features}

\begin{figure*}
\centering
\includegraphics[width=0.849\textwidth]{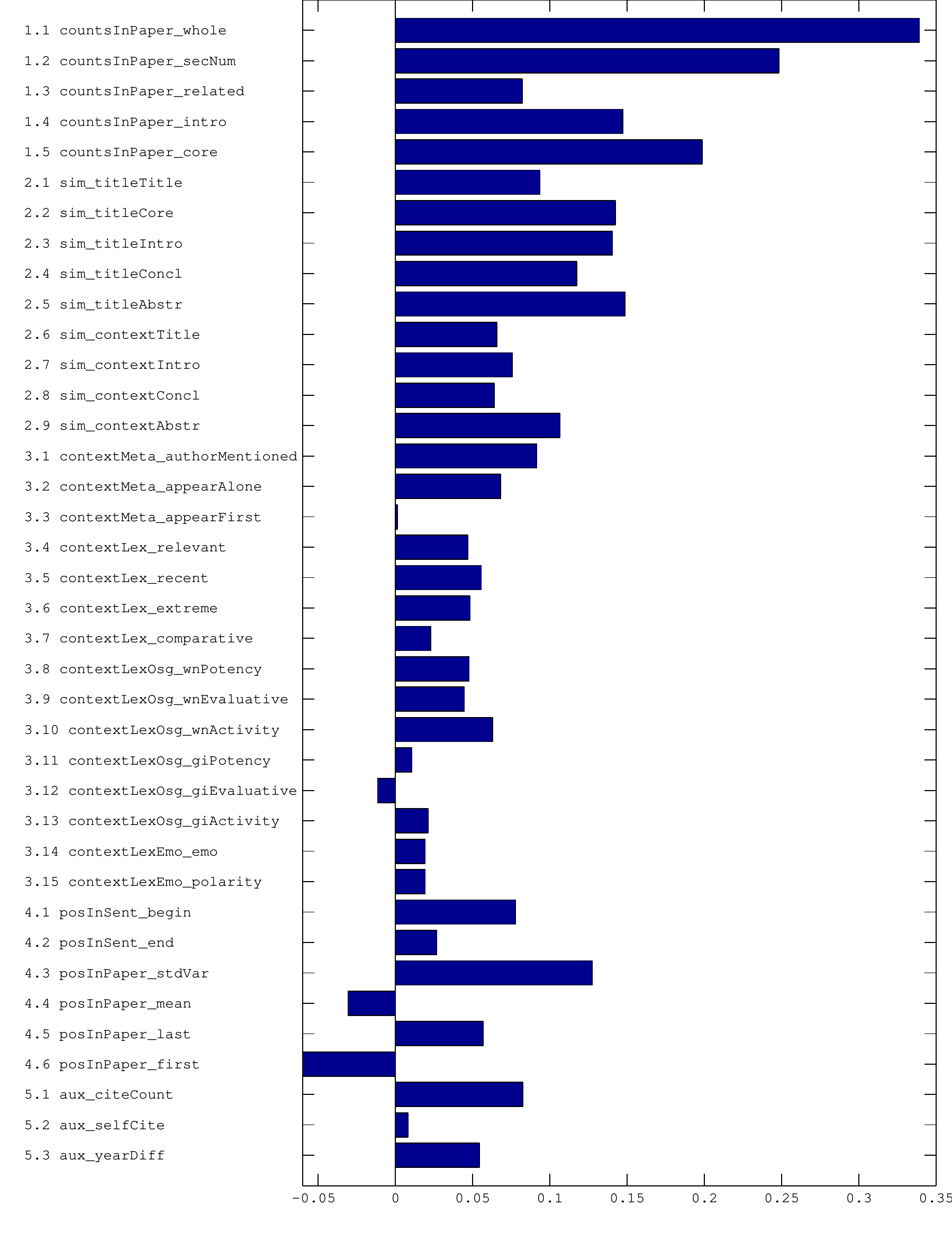}
\caption{Pearson correlation coefficients between the features and the
  gold influence labels. The y-axis lists the features and
  x-axis the coefficients.  }
\label{fig:correlation}
\end{figure*}

We seek to determine which features are better able to predict the
academic influence of a reference.  The Pearson correlation
coefficients between the various features and the gold influence
labels are a simple indication of how useful a feature might be.
We show the coefficients in  Figure~\ref{fig:correlation}.

First, consider the correlation coefficients for the count-based
features:

\begin{enumerate}[itemsep=1pt,parsep=1pt,topsep=1pt,partopsep=1pt]

\item Count-based features 

\begin{enumerate}[itemsep=1pt,parsep=1pt,topsep=1pt,partopsep=1pt,label*=\arabic*.]

\item countsInPaper\_whole
\item countsInPaper\_secNum
\item countsInPaper\_related
\item countsInPaper\_intro
\item countsInPaper\_core

\end{enumerate}
\end{enumerate}

\noindent 

Figure~\ref{fig:correlation} shows that the most correlated individual 
features  to academic influence are in-paper count features 
($counts\allowbreak{}InPaper\allowbreak{}\_whole$ and 
$counts\allowbreak{}InPaper\allowbreak{}\_secNum$). This is a convenient
result, because one of the best features, $counts\allowbreak{}InPaper\allowbreak{}\_whole$ 
(the number of times a reference is cited in a paper), is also one of the easiest to compute 
from a technical point of view. Moreover, it suggests simple but potentially 
effective schemes for modifying the standard citation count (i.e., the
number of papers that cite a given paper in the general literature):
For each paper $X$ that cites a paper $Y$, increment the citation count
for $Y$

\begin{itemize}[itemsep=1pt,parsep=1pt,topsep=1pt,partopsep=1pt]

\item only if $Y$ was cited more than once in the body of $X$, 

\item only if $Y$ is cited more often in the body of $X$ than most of the 
other references in $X$, or 

\item only if $Y$ is cited in more than one section of $X$. 

\end{itemize}

\noindent Intuitively, these modified citation counts may also be more robust than 
the standard citation count, considering that an author seems unlikely to 
cite a paper more than once when the paper is included in the references
because it is \emph{de rigueur} in the field.

Next, consider the similarity-based features. These are features
that measure the semantic similarity between a citer and a reference.
In general, we found such features well correlated with academic influence. 

The first group of similarity-based features compares the similarities between 
the title of a cited paper and the title, introduction, conclusion, and abstract 
of the citer:

\begin{enumerate}[itemsep=1pt,parsep=1pt,topsep=1pt,partopsep=1pt,resume]

\item Similarity-based features

\begin{enumerate}[itemsep=1pt,parsep=1pt,topsep=1pt,partopsep=1pt,label*=\arabic*.]

\item sim\_titleTitle
\item sim\_titleCore
\item sim\_titleIntro
\item sim\_titleConcl
\item sim\_titleAbstr

\end{enumerate}
\end{enumerate}

\noindent As shown in Figure~\ref{fig:correlation}, the correlation coefficients of the 
features (e.g., $sim\allowbreak{}\_title\allowbreak{}Abstr$) rank right after those of 
the two in-paper count features. As we will show soon, these features (e.g., $sim\_titleCore$) can
work synergetically with count-based features for predicting academic influence.

The second group of similarity-based features use citation context
instead of the title of a cited paper:

\begin{enumerate}[itemsep=1pt,parsep=1pt,topsep=1pt,partopsep=1pt,resume,start=2]

\item Similarity-based features

\begin{enumerate}[itemsep=1pt,parsep=1pt,topsep=1pt,partopsep=1pt,resume,start=6,label*=\arabic*.]

\item sim\_contextTitle
\item sim\_contextIntro
\item sim\_contextConcl
\item sim\_contextAbstr

\end{enumerate}
\end{enumerate}

\noindent These features compare the similarities between citation contexts 
and the title, abstract, and conclusion of the citing paper. We found that the 
context--abstract ($sim\_contextAbstr$) similarity feature is the one in this group
that is most correlated 
with academic influence, followed by context--conclusion ($sim\_contextIntro$),
context--title($sim\_contextTitle$), and context--introduction
($sim\_contextConcl$).

We turn to the context-based features. First, we focus on the features that
consider the relation between the citation and the citation context:

\begin{enumerate}[itemsep=1pt,parsep=1pt,topsep=1pt,partopsep=1pt,resume]

\item Context-based features

\begin{enumerate}[itemsep=1pt,parsep=1pt,topsep=1pt,partopsep=1pt,label*=\arabic*.]

\item contextMeta\_authorMentioned
\item contextMeta\_appearAlone
\item contextMeta\_appearFirst

\end{enumerate}
\end{enumerate}

\noindent In this group, $contextMeta\_authorMentioned$ has the highest correlation
coefficient. This feature indicates whether the names of the authors appear
in the citation context (e.g., ``Smith et al. [4]''). 

The second group of context-based features are based on the meaning of the 
words in the citation context:

\begin{enumerate}[itemsep=1pt,parsep=1pt,topsep=1pt,partopsep=1pt,resume,start=3]

\item Context-based features

\begin{enumerate}[itemsep=1pt,parsep=1pt,topsep=1pt,partopsep=1pt,resume,start=4,label*=\arabic*.]

\item contextLex\_relevant
\item contextLex\_recent
\item contextLex\_extreme
\item contextLex\_comparative
\item contextLexOsg\_wnPotency
\item contextLexOsg\_wnEvaluative
\item contextLexOsg\_wnActivity
\item contextLexOsg\_giPotency
\item contextLexOsg\_giEvaluative
\item contextLexOsg\_giActivity
\item contextLexEmo\_emo
\item contextLexEmo\_polarity

\end{enumerate}
\end{enumerate}

\noindent We used several different types of lexicons to capture different aspects of semantics 
in the citation contexts, including sentiment, emotion, and~\citeS{Osgood:1957} semantic
differential categories. 

As we mentioned in the preceding section, $contextLexEmo\_polarity$ is the number of
words in the citation context that are labeled either positive or negative and
$contextLexEmo\_emo$ is the number of words that are labeled with any of the eight 
basic emotions. Our hope was that any kind of sentiment or emotion in the words in the 
citation context might indicate that the citation is influential, even if the sentiment or emotion
is negative, but Figure~\ref{fig:correlation} shows that neither feature has
a high correlation with the gold labels. This suggests to us that it might be
better to split these features into more specific features for each of the possible
categories. 

\begin{figure}
\centering
\includegraphics{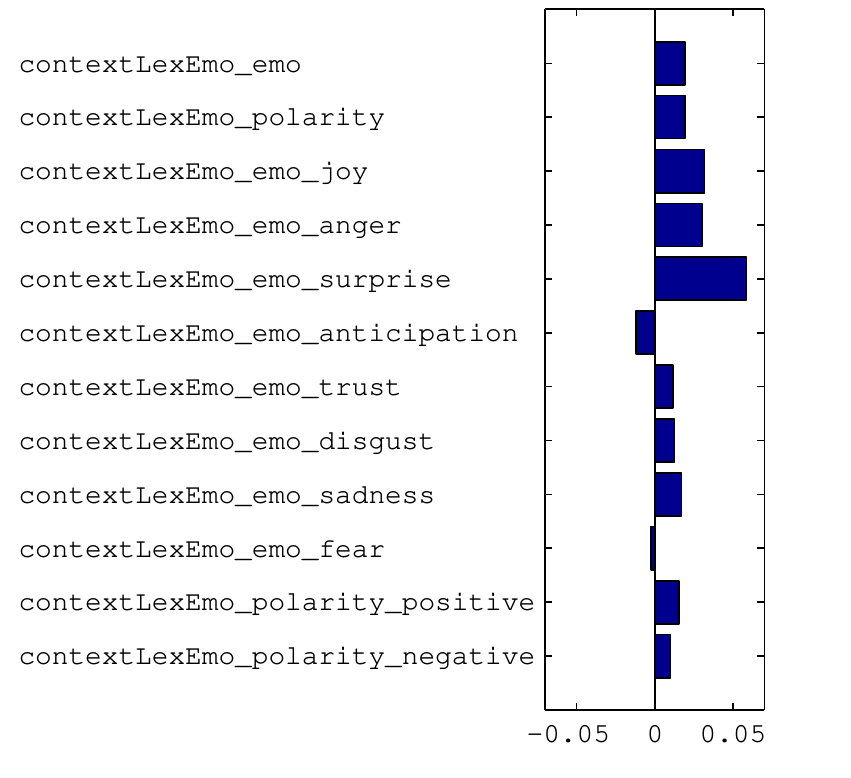}
\caption{Detailed Pearson correlation coefficients between emotional and 
sentimental features and the gold labels.}
\label{fig:emo_correlation}
\end{figure}

To test this idea, we split $contextLexEmo\_polarity$ into two features,
one for positive polarity and one for negative polarity, and we split 
$contextLexEmo\_emo$ into eight features, one for each of the eight basic emotions.
Figure~\ref{fig:emo_correlation} shows the correlation coefficients for each of
these more specific features.

We see in Figure~\ref{fig:emo_correlation} that positive polarity has a higher
correlation than negative polarity. Among the eight basic emotions, surprise has
the highest correlation. These results are intuitively reasonable. However, none
of the correlations is greater than 0.06. It seems that none of these features
are likely to be of much use for predicting influence.

Let us consider the position-based features. First, we examine the position
of a citation in a sentence:

\begin{enumerate}[itemsep=1pt,parsep=1pt,topsep=1pt,partopsep=1pt,resume]

\item Position-based features

\begin{enumerate}[itemsep=1pt,parsep=1pt,topsep=1pt,partopsep=1pt,label*=\arabic*.]

\item posInSent\_begin
\item posInSent\_end

\end{enumerate}
\end{enumerate}

\noindent Our results suggest that references located at the beginning of 
a sentence might be more influential.

The second group of position-based features calculates the locations of
citations in the body of the paper:

\begin{enumerate}[itemsep=1pt,parsep=1pt,topsep=1pt,partopsep=1pt,resume,start=4]

\item Position-based features

\begin{enumerate}[itemsep=1pt,parsep=1pt,topsep=1pt,partopsep=1pt,resume,start=3,label*=\arabic*.]

\item posInPaper\_stdVar
\item posInPaper\_mean
\item posInPaper\_last
\item posInPaper\_first

\end{enumerate}
\end{enumerate}

\noindent The best paper-position-based feature is the standard variance of a 
reference's positions ($posInPaper\_stdVar$). Note that this feature is likely to
overlap with the in-paper-counts features to some degree: A larger in-paper-counts 
number could correspond to a higher position variance, so these features may not 
have additive benefit when used together.

Finally, we examine the miscellaneous features:

\begin{enumerate}[itemsep=1pt,parsep=1pt,topsep=1pt,partopsep=1pt,resume]

\item Miscellaneous features

\begin{enumerate}[itemsep=1pt,parsep=1pt,topsep=1pt,partopsep=1pt,label*=\arabic*.]

\item aux\_citeCount
\item aux\_selfCite
\item aux\_yearDiff

\end{enumerate}
\end{enumerate}

\noindent Figure~\ref{fig:correlation} shows that the correlation coefficient between 
citation counts and the influence labels is positive. This confirms the previous finding that
highly cited papers are more likely to be cited in a meaningful manner~\cite{Bornmann2008}.
However, we find that the correlation is moderate: It is smaller than that of half of 
the features we tested. Hence, while highly cited papers may have more academic influence, 
the citation count is not an ideal indicator of influence. This result is consistent 
with the fact that papers are often cited for reasons other than academic influence.
When a paper is highly cited, the authors of the citing
paper may feel obliged to cite it, yet might not take the
time to read it.

From Figure~\ref{fig:correlation}, we see a small positive correlation between self-citation 
and the gold influence labels. We will see later that $aux\_selfCite$ is useful
as a corrective factor in the final model.

\begin{figure}
\centering
\includegraphics[trim=1.15cm 0.1cm 0.5cm 0.6cm, clip=true, width=0.49\textwidth]{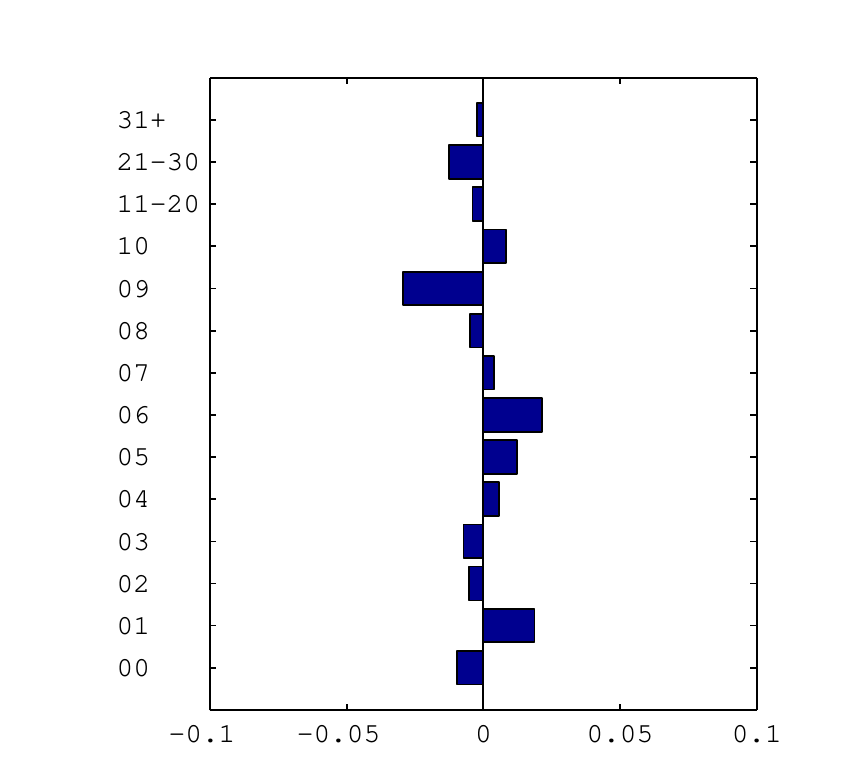}
\caption{Detailed Pearson correlation coefficients between aux\_yearDiff and the gold
  labels. }
\label{fig:yearDiff}
\end{figure}

Are older papers more likely to be influential? In Figure~\ref{fig:correlation}, the 
$aux\_yearDiff$ feature has a small positive correlation with the influence. We 
discretized the feature over the ranges 0, 1, \ldots, 10, 11--20, 21--30, $31+$.
The corresponding Pearson coefficients are given in Figure~\ref{fig:yearDiff}.
The figure shows that, when the cited paper is one or four to seven years older than the citing 
paper, there is a positive correlation with academic influence. More recent papers 
(0, 2 and 3~years) and older papers ($\geq 8$~years) are poorly or negatively correlated 
with academic influence. This result is consistent with our interest (mentioned
in the section on defining academic influence) in the \emph{proximate} influences
on a citing paper.

\subsection*{Results of predicting academic influence}

We used the LIBSVM support vector machine (SVM) package~\cite{Chang:2011} 
as our supervised learning algorithm.\footnote{LIBSVM is available for
download at \url{http://www.csie.ntu.edu.tw/~cjlin/libsvm/}.} We chose a 
second-degree polynomial as our kernel function.\footnote{The LIBSVM 
parameters we used are ``-s~0~-d~2~-t~1~-r~1''.} 

The \mbox{F-measure} was applied to evaluate the performance of the learned model. 
The \mbox{F-measure} is defined as the harmonic mean of precision $P$ and recall~$R$: 
$F = 2PR / (P+R)$. The precision of a model is the conditional probability that a 
paper--reference pair is \emph{influential} (according to the gold-standard author-generated 
label), given that the model guesses that it is \emph{influential}. The recall of a model
is the conditional probability that the model guesses that a paper--reference pair is 
\emph{influential}, given that it actually is \emph{influential} (according to the gold-standard).  

Another metric is \emph{accuracy}, defined as the ratio of properly classified 
paper--reference pairs (as either \emph{influential} or not) over the total number of classified pairs.
However, the classes in our data are imbalanced (10.3\% in the \emph{influential} class and 
89.7\% in the \emph{non-influential} class). This makes accuracy inappropriate as a 
performance measure for our task, because we could achieve an accuracy of 89.7\% by the 
trivial strategy of always guessing the \emph{non-influential} class. 

The \mbox{F-measure} is a better performance measure for imbalanced classes, because
it rewards a model that has a balance of precision and recall. Always guessing 
\emph{non-influential} yields an \mbox{F-measure} of zero (we take division by zero to
be zero). Always guessing \emph{influential} yields an \mbox{F-measure} of 18.7\%
($2 \cdot 0.103 \cdot 1 / (0.103 + 1) = 0.187$). Unlike accuracy, the \mbox{F-measure}
penalizes trivial models.

The SVM algorithm is designed to optimize accuracy, whereas we want to optimize
the \mbox{F-measure}. Since an SVM does not directly optimize the \mbox{F-measure} and our data 
are not balanced, we used a simple down-sampling method to handle this. In each fold of 
cross validation, we randomly down-sampled the negative instances (non-influential references) 
in the training data to make their number equal to that of the positive
ones (influential references).

Table~\ref{tab:bestSys} shows the scores of different models under ten-fold
cross-validation. We included two baselines. The first baseline randomly
labels a reference with a probability equal to the distribution of the
labels in the training data. The second predicts the academic influence of
references based on their Google citation counts ($aux\_citeCount$).
Note that the macro-averaged F-measure is not necessarily between the averaged precision and recall. For example, for model~(3) in the table, 0.35 is not between 0.36 and 0.41.

Starting with model~(3) in the table, we added features greedily: In each
round, the feature resulting in the maximum improvement of the \mbox{F-measure} 
was added. That is, model~(3) is the best model that uses only one single 
feature, and the best performing model (with four features) is model~(6).

\begin{table}
\caption{Additive improvement of model performance. We report macro-averages under ten-fold
cross-validation. 
With respect to the \mbox{F-measure}, asterisks (*) mark models that are better than model (2) whereas
daggers (\dag) mark models that are better than model (3).}
\label{tab:bestSys}
\centering
\begin{tabular}{clSlSS}
\hline
Model & Features &\multicolumn{2}{c}{F-measure} & {Precision} & {Recall}\\
\hline
(1)   & random                       & 0.10 &       & 0.10 & 0.10 \\
(2)   & aux\_citeCount               & 0.12 &       & 0.12 & 0.13 \\
\hline
(3)   & countsInPaper\_whole         & 0.35 & *     & 0.36 & 0.41 \\
(4)   & (3) + sim\_titleCore         & 0.39 & *\dag & 0.40 & 0.44 \\
(5)   & (4) + countsInPaper\_secNum  & 0.41 & *\dag & 0.42 & 0.46 \\
(6)   & (5) + aux\_selfCite          & 0.42 & *\dag & 0.43 & 0.48 \\
\hline
\end{tabular}
\end{table}

In Table~\ref{tab:bestSys}, all of the models marked with a dagger sign 
(${\dag}$) are statistically significantly better than model~(3). The 
models marked with an asterisk (*) are statistically significantly better than the 
two baselines. We use a one-tailed paired t-test with a 99\% significance
level.

The first feature chosen for the model by greedy feature selection is
$counts\allowbreak{}InPaper\allowbreak{}\_whole$, the feature with the
highest correlation in Figure~\ref{fig:correlation}. The best model
achieves an \mbox{F-measure} of about 42\% (see Table~\ref{tab:bestSys}).
The model uses only four features, two of which are count-based
($countsInPaper$) and one semantics-based ($sim\_titleCore$). Adding more 
features to model (6) did not result in further improvement. Using all features
presented in Figure~\ref{fig:correlation} results in an \mbox{F-measure} of
37\%, which is significantly better than model (1) and (2) ($p<0.01$),
insignificantly better than the best single-feature model (3)
($p>0.05$), and worse than model (6) ($p<0.01$). This observation
supports the hypothesis that feature selection is useful for this task. In
general, feature selection removes useless or detrimental features, which 
often leads to better performance (e.g., higher \mbox{F-measures}) and greater 
efficiency.

We find it interesting that a semantic feature ($sim\_titleCore$, the
similarity between the title of the cited paper and the core sections
of the citing paper) is the second feature chosen. It seems that this
feature complements the count-based feature; it covers some papers
that are influential but have lower counts. The improvement of model
(4) over model (3) is about 4\% in terms of \mbox{F-measure}, which is
statistically significant at a level of $p<0.01$. Using another
count-based feature, $counts\allowbreak{}InPaper\allowbreak{}\_secNum$,
additionally improves the performance. Although $aux\_selfCite$ by
itself has a small correlation with influence (see Figure~\ref{fig:correlation}), 
it seems to be useful when combined with the other three features.

Note that the \mbox{F-measure} we used here is the macro-averaged
\mbox{F-measure} \cite{Lewis:1991}. That is, we calculated
the \mbox{F-measure} for each paper individually and then computed the arithmetic
average over all the \mbox{F-measures} obtained. For each
reference in a given paper, we used the SVM model to estimate
the probability that the reference is labeled {\em influential}.
In the training data, the average paper contained three influential references.
Therefore the model guesses that the top three references in the
given paper, with the highest estimated probabilities, are {\em influential},
and the remaining references, with lower probabilities, are {\em non-influential}.

It is difficult to describe an SVM model intuitively. Moreover, 
given an SVM model, it is not straightforward to describe the importance
of a feature. To further assess the importance of the four features in
model~(6), we have also applied logistic regression to our data~\cite{long1997regression}.
Logistic regression assumes that the probability distribution of some binary 
random variable $Y$ is of the form 

\begin{equation*}
P(Y=1|\vec{X})= \frac{1}{1+e^{-(\beta_0+\sum_i \beta_i X_i)}}
\end{equation*}

\noindent where  $\vec{X}=(X_1, X_2, \ldots )$ are feature values and
$\beta_0, \beta_1, \beta_2, \ldots$ are weight vectors. Given $N$ 
training  instances, we can solve for the weights 
$\vec{W} = (\beta_0, \beta_1, \ldots)$ by maximizing the likelihood function

\begin{equation*}
l(\vec{W}) = \prod\nolimits_{i=1}^N P(Y=1|X)^{t_i}(1-P(Y=1|X))^{1-t_i}
\end{equation*}

\noindent where $t_i \in \left\{0,1\right\}$ is the binary gold label of the $i^{th}$ training 
instance. Once we have solved for the weights, we can classify instances by using 
a threshold $\omega$. That is, given an instance with feature values $\vec{X}$, we
predict $Y=1$ if $P(Y=1|\vec{X})>\omega$ and we predict $Y=0$ otherwise. For our 
application, we set the threshold so that the relative number of influential 
citations is the same as in the training set.

With logistic regression, the magnitude (absolute value) of the weights, $\beta_i$,
indicates the importance of the corresponding feature in the model.
We used the \textit{mnrfit} command in Matlab to conduct logistic
regression on our data. The weights assigned to the features
$countsInPaper\_whole$, $sim\_titleCore$, $countsInPaper\_secNum$, and
$aux\_selfCite$ are 2.7228, 1.2683, 1.1763, and -0.0923. Their absolute 
values correspond to the order they are selected by SVM in Table~\ref{tab:bestSys}: 
The most important feature is $countsInPaper\_whole$ and the least important is 
$aux\_selfCite$. Note that the weight for $aux\_selfCite$ is smallest, which 
corresponds to the observation that self-citations are less likely to be 
influential. We have also used logistic regression to classify
the references, using the same experimental setup as we used for the SVM\@.
The logistic regression performance is slightly below that of the SVM 
($\approx 0.37$ vs. $0.41$).\footnote{With logistic regression, 
the precision and recall are 0.38 and 0.42.}

\begin{figure}
\centering
\includegraphics[clip=true, width=0.7\textwidth]{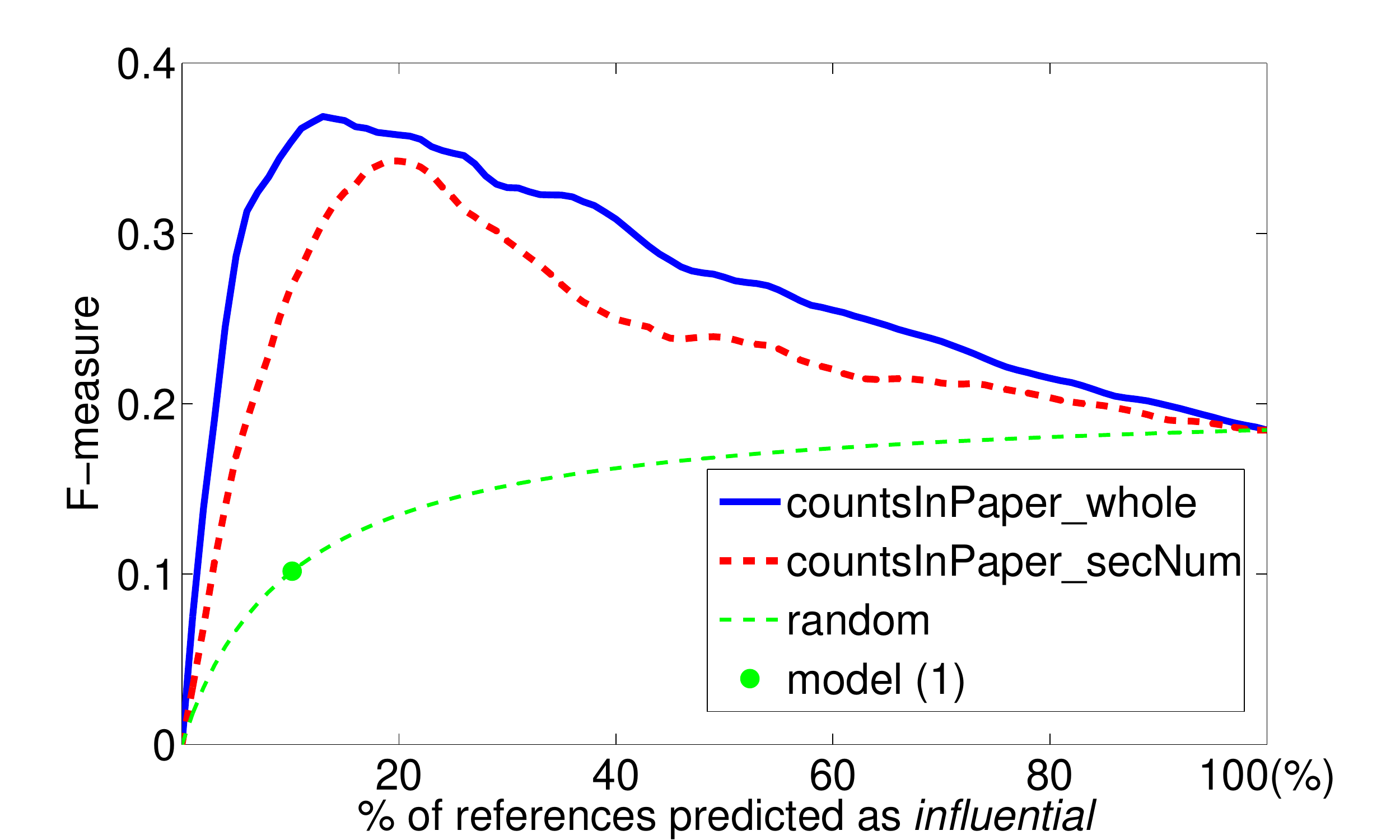}
\caption{Detailed \mbox{F-measure} curves for \emph{countsInPaper\_secNum} 
and \emph{countsInPaper\_whole}.}
\label{fig:threshClr}
\end{figure}

When only one feature is used with SVM (as in model (3)), the classification
task can be regarded as setting a threshold on the feature, to
separate the influential references from the rest. In
Figure~\ref{fig:threshClr}, we vary such a threshold to provide a full
view of the \mbox{F-measures} of the two most relevant features shown in
Figure~\ref{fig:correlation}. Different thresholds here resulted in
different percentages of references being predicted
as \emph{influential}, corresponding to the x-axis in the figure. Note
that our thresholds are in the range [0, 1], since we have normalized
the count values (and all other features) to this range (as we
discussed earlier).

Figure~\ref{fig:threshClr} also includes the \mbox{F-measure} 
curve of random guesses (\emph{random} in the figure). This curve
serves as a minimum baseline for comparison with the other features.
Model (1) in Table~\ref{tab:bestSys} corresponds to the point
on the random curve such that the percentage of references predicted
as \emph{influential} equals the size of the \emph{influential}
class ( 322/3143 = 10.3\%).

In Figure~\ref{fig:threshClr}, the peak \mbox{F-measure} of our best feature,
$countsInPaper\_whole$, is 0.37  (when the value on the x-axis is 13\%).
Model~(3)  attained an \mbox{F-measure} of 0.35 in Table~\ref{tab:bestSys},
slightly below the value of 0.37 in Figure~\ref{fig:threshClr}.
Model (3) was trained and tested with ten-fold cross validation,
whereas the \mbox{F-measure} of 0.37 is based on using the whole dataset
as training data, with no independent testing data; thus the small gap
in the \mbox{F-measures} (0.35 versus 0.37) indicates that SVM is performing well.

\section*{Experiments with in-paper citation counts}

In the preceding section, we made a number of observations. A
significant one is that the in-paper citation counts (how many
times a reference is cited in a paper) are the most predictive
features for academic influence. The following experiments are
designed to further validate our results. Since the in-paper counts 
convey influence information, which is ignored
in the conventional counting of citations, we wondered whether
incorporating in-paper citation numbers into global citation counting
would result in different rankings of papers and authors.

In contrast to the conventional citation counting, we refer to the methods
that take into consideration the in-paper counts as \emph{influence-primed citation
counts}. We conducted two types of experiments. First,
we explored the correlation between the rankings of papers and authors
with and without influence-primed citation counts. Second, we tackled 
the task of identifying ACL Fellows.

\subsection*{Influence-primed citation counts}

A classical citation network is a graph in which the nodes (vertices)
correspond to papers and there is a directed link (directed edge)
from one paper to another if the first paper cites the second
paper. The network is usually acyclic (it has no loops), due to
time: An earlier paper rarely cites a later paper (with some
exceptions, due to overlap in the gestation periods of publications).

A slightly more sophisticated citation network could have weights
or labels associated with the edges in the graph. For example,
a directed edge from citing paper $X$ to cited paper $Y$ might be 
labeled as \emph{$Y$ provides evidence that supports claims in $X$},
or the directed edge might be weighted with a number that indicates
how influential $Y$ is to $X$.

There are various ways that in-paper citation counts can be used
to modify classical citation networks. We have experimented with
using in-paper citation counts for \emph{filtering} edges in the
graph and for \emph{weighting} edges in the graph.

A simple filtering method is to drop the edge from citing paper $X$ 
to cited paper $Y$ when the in-paper citation count for $Y$ in $X$
is below a threshold. Equivalently, when building a citation network,
only add an edge when the in-paper citation count is greater than
or equal to a threshold.

We have also tried filtering with a combination of two thresholds, $T_1$ 
and $T_2$. For a given paper--reference pair (i.e., a given citing--citer 
pair in the citation network), when building a network, we add an edge 
from citing paper $X$ to cited paper $Y$ based on the in-paper citation 
count of the reference $Y$ (i.e., the number of times $Y$ is mentioned 
anywhere in the body of $X$) and the rank of the reference $Y$ relative to 
the other references in $X$ (i.e., the rank in a list, sorted in descending
order of in-paper citation counts). An edge is added only if
the in-paper citation count is at least $T_1$ and the rank is less than
$T_2$ (lower rank is better, because the list is sorted in descending order).

An alternative to filtering is weighting edges. The edge between a reference 
and a citer can be weighted by the in-paper citation counts. Given a citing paper 
$X$ and a cited paper $Y$, suppose that $Y$ is mentioned $c$ times in the body of $X$. 
There are many functions that we might apply to convert the in-paper citation count 
$c$ to a weight. Any linear or polynomial function might be useful.

In the following experiments, we use the square of the in-paper count to weight an 
edge: We weight the edge from $X$ to $Y$ with $c^2$. Squaring $c$ gives more 
weight to higher values of $c$. 

The \emph{conventional citation count} for a paper is calculated from a citation network.
It is the number of edges in the graph that are directed into the vertex that represents
the given paper. That is, the conventional citation count for a paper is the number
of papers that cite the given paper.

We weight citations according to the square of the number of times the reference
is mentioned in the text. For example, being cited once in a paper that mentions the 
reference once counts for one whereas being cited once in a paper that mentions the 
reference twice counts for four. Weights are added up: Being cited four times by four papers 
that mention the reference once counts the same as being cited once by a paper that 
mentions the reference two times. 

The \emph{influence-primed citation count} is like the conventional citation count,
except it weights each edge by $c^2$, instead of 1. We define \emph{influence-primed 
citation count} formally as follows:

\begin{definition}
Given two papers, $p_i$ and $p_j$, let $c(p_i,p_j)$ be the number of times paper 
$p_i$ mentions paper $p_j$ in the body of its text, excluding the reference section. 
If the paper $p_i$ cites paper $p_j$, then $c(p_i,p_j) > 0$; otherwise $c(p_i,p_j) = 0$. 
Let $L$ (the literature) be the set of all papers in the given citation network. 
The \emph{influence-primed citation count} of paper $p_j$, $\textrm{cip}(p_j)$, is
\begin{equation*}
\sum_{p_i \in L} c(p_i,p_j)^2.
\end{equation*}
\end{definition}

\noindent The function name, $\textrm{cip}(\cdot)$, stands for \emph{citations, influence-primed}.

The \emph{conventional \mbox{$h$-index}} for an author is the largest number $h$
such that at least $h$ of the author's papers are cited by at least $h$ other papers. 
Each citation of a paper has a weight of 1.

The \emph{influence-primed \mbox{$h$-index}} for an author is like the conventional
\mbox{$h$-index}, except it weights each edge by $c^2$, instead of 1. 
The \emph{influence-primed \mbox{$h$-index}} for an author is
the largest number $h$ such that at least $h$ of the author's papers have an
\emph{influence-primed citation count} of at least $h$.

For example, if an author has four papers, each one cited only once, but each time they
are cited, they are mentioned twice, then the influence-primed \mbox{$h$-index} is 4. In
contrast, with the conventional \mbox{$h$-index}, the same author would receive an \mbox{$h$-index}
of 1. 

We define \emph{influence-primed \mbox{$h$-index}} formally as follows:

\begin{definition}
An author, $a_i$, with a set of papers $O(a_i)$ (the \oe{}uvre) has an 
\emph{influence-primed \mbox{$h$-index}}, $\textrm{hip}(a_i)$, of $h$ if $h$ 
is the largest value such that
\begin{equation*}
| \{p_j \in O(a_i) | \textrm{cip}(p_j)\geq h\} | \geq h.
\end{equation*}
\end{definition}

\noindent The function name, $\textrm{hip}(\cdot)$, stands for \emph{\mbox{$h$-index}, influence-primed}.
We refer to this as the hip-index.

\subsection*{ACL Anthology Network}

For this experiment, we use the AAN (ACL Anthology Network) dataset~\cite{radev2009acl}. 
AAN is a citation network constructed from the papers published in Association for 
Computational Linguistics (ACL) venues (conferences, workshops, and journals since 1965),
approximately 20,000 papers. The AAN citation network is a closed graph; edges from or to the papers 
published outside ACL venues are not included. 
In effect, by using the AAN citation network, we can measure the  impact of researchers 
and papers on the ACL community. This restriction might be desirable when we try to 
identify the recipients of honours granted by ACL\@. 
Table~\ref{tab:aan} shows the basic statistics for the 
dataset.\footnote{The AAN corpus is available at \url{http://clair.eecs.umich.edu/aan/index.php}.}

\begin{table}
\caption{Statistics of the AAN dataset.}
\label{tab:aan}
\centering
\begin{tabular}{lr}
\hline
Statistics                & Values\\
\hline
Number of venues          &    341\\
Number of papers          & 18,290\\
Number of authors         & 14,799\\
Number of paper citations & 84,237\\
\hline
\end{tabular}
\end{table}

To obtain in-paper counts, we used regular expressions to locate citations. Our
regular expressions are good at both precision and recall according to our manual examination, 
but they still make a few errors. For example, the regular expressions have trouble with
citations that span two lines of text and with multiple papers written by the same author 
in the same year. Another problem is automatically distinguishing the main body text
from the reference section of a paper. The regular expressions may wrongly increment
the in-paper count by matching citations in the reference section.

Numerical citations (e.g., ``[1]'' or ``[1,2,3]'') are more difficult to process
than textual citations (e.g., ``Smith et al. (1998)''). We used a random number generator 
to select a sample of 100 papers from AAN and then we manually determined their citation types. 
In this random sample, 7\% used numerical citations. Since numerical citations
are relatively rare in the AAN dataset, we simply ignored them.

We did not normalize the in-paper citation count $c$ for these experiments with
the AAN dataset. In this section, the in-paper citation count $c$ is a non-negative 
integer value. The reason for this is that the AAN network is a closed graph: All citations
to and from papers outside of the AAN dataset are ignored in the AAN citation network.
The maximum value that we used for contextual normalization in the preceding section, 
$\max (p_i,r_{i*},f_k)$, could be distorted by the ignored citations.
For example, a paper that mainly cites the AAN papers could be normalized very differently 
from one that mainly cites non-AAN papers. The maximum value may be highly sensitive to 
whether a citing paper is influenced by cited work that is outside of the AAN network.

The in-paper citation count $c$ that we use in this section is essentially a raw 
(unnormalized) variation of $countsInPaper\_whole$. We did not use 
$counts\allowbreak{}InPaper\allowbreak{}\_secNum$,
because it might be more sensitive to noise introduced by the process of automatically
detecting section boundaries. (We manually detected sections in the preceding experiments,
but this manual process does not scale up from 100 papers to 20,000 papers.)
Figure~\ref{fig:threshClr} suggests that $countsInPaper\_whole$ performs better than
$countsInPaper\_secNum$, and it is easier to compute.

\subsection*{Conventional versus influence-primed counting}

A natural question is whether there is any difference
between conventional citation counts and influence-primed citation
counts. In particular, do the two approaches yield different rankings
of the papers?

Table~\ref{tab:paperRank} shows the Spearman correlation coefficients between 
the AAN papers. We grouped the papers according to their ranks in the conventional
counting. For each group, we calculated the Spearman correlation coefficient between 
the conventional counts and the influence-primed counts.

\begin{table}
\caption{Spearman correlation between conventional and influence-primed
citation counts for groups of AAN papers.}
\label{tab:paperRank}
\centering
\begin{tabular}{rS}
\hline
Papers    & {Correlations}\\
\hline
1--100    &  0.67 \\
101--200  &  0.12\\
201--300  &  0.11\\
301--400  & -0.04\\
401--500  &  0.07\\
501--600  &  0.05\\
601--700  & -0.13\\
701--800  &  0.30\\
801--900  &  0.22\\
901--1000 &  0.06\\
\hline
\end{tabular}
\end{table}

For example, papers 1--100 are the top 100 most highly cited papers, according
to conventional citation counts, where each edge directed into a given paper
increments that paper's count by one. For these 100 papers, we have a vector
of 100 conventional citation counts. We also calculate a vector of 100 
influence-primed citation counts. The Spearman correlation between these two vectors
is 0.67.\footnote{Spearman correlation is specifically intended for comparing
ranked lists, whereas Pearson correlation is more appropriate when numerical
values are more important than ranks.}

For the top 100 most highly cited papers, conventional citations counts
and influence-primed citation counts have a high correlation. As we move
down the list, the correlation drops. The two counts agree on the most
highly ranked papers, but they disagree on the less cited papers.
Weighting makes a difference.

Table~\ref{tab:authorRank} shows the Spearman correlation coefficients for 
the AAN authors, under these two different counting methods.
The \mbox{$h$-indexes} of the authors were calculated
and were used to rank the authors. For each group of authors, we
calculate the Spearman correlation between the \mbox{$h$-indexes}
and the hip-indexes. Comparing Tables~\ref{tab:paperRank} and~\ref{tab:authorRank}, 
we see that the authors' correlations steadily decline as we go down the rows
of Table~\ref{tab:authorRank}, but the papers' correlations fluctuate 
with no clear trend as we go down the rows of Table~\ref{tab:paperRank}.
This indicates that conventional citation counts and influence-primed citation 
counts are not related by a simple linear transformation. That is, influence-primed
counting is different from conventional counting in a non-trivial way.

\begin{table}
\caption{Spearman correlation between conventional \mbox{$h$-indexes} and 
 hip-indexes for groups of AAN authors.}
\label{tab:authorRank}
\centering
\begin{tabular}{rS}
\hline
Authors   & {Correlations}\\
\hline
1--100    &  0.74\\
101--200  &  0.49\\
201--300  &  0.21\\
301--400  &  0.40\\
401--500  &  0.22\\
501--600  &  0.10\\
601--700  &  0.01\\
701--800  &  0.03\\
801--900  & -0.04\\
901--1000 &  0.03\\
\hline
\end{tabular}
\end{table}

\subsection*{Identifying ACL Fellows}

The Association for Computational Linguistics has seventeen fellows.\footnote{ACL 
Fellows are listed at \url{http://aclweb.org/aclwiki/index.php?title=ACL_Fellows}.
When we performed our experiments, there were only seventeen fellows, but there
are more now.} We might assume that these seventeen fellows can be identified
by selecting the authors having the best \mbox{$h$-index} scores. Indeed, 
\citeA{hirsch2007does} found that the \mbox{$h$-index} was superior 
at predicting the future performance of a researcher than the
number of citations, the number of papers, and mean citations per 
paper~\cite{lehmann2006measures}.

Table~\ref{tab:fellow} shows the precision of the conventional \mbox{$h$-index}
and influence-primed hip-index at identifying ACL Fellows. For a given value of
$N$, we sort all authors in AAN in descending order of their \mbox{$h$-indexes}
and then count the number of ACL Fellows among the top $N$ authors in the sorted
list. We also sort all authors in AAN in descending order of the hip-indexes
and then count the number of ACL Fellows among the top $N$. Precision is the number 
of ACL Fellows in the top $N$ divided by $N$. Table~\ref{tab:fellow} shows the precision
as $N$ ranges from 1 to 17: we stop at 17 because there are 17~ACL~Fellows in total. 
For example, the second row in the body of the table, with 
$N = 2$, shows that zero of the top two \mbox{$h$-index} ranked authors are ACL Fellows 
(precision 0\%), but one of the top two hip-index ranked authors is an ACL Fellow
(precision 50\%). For $N = 3$, \mbox{$h$-index} finds zero ACL Fellows but hip-index 
finds two Fellows.

\begin{table}[th]
\caption{Precisions of the $h$-index and  of the hip-index at 
identifying ACL~fellows. Cases where the precisions differ are indicated with asterisks. 
In all such cases, the influence-primed model is more precise.}
\label{tab:fellow}
\centering
\begin{tabular}{SSSS}
\hline
       & \multicolumn{3}{c}{Precision (\%)} \\
{$N$}  & {$h$-index} &   & {hip-index} \\
\hline
1      &  0          &   & 0  \\
2      &  0          & * & 50 \\
3      &  0          & * & 67 \\
4      & 25          & * & 50 \\
5      & 20          & * & 40 \\
6      & 33          &   & 33 \\
7      & 29          &   & 29 \\
8      & 25          &   & 25 \\
9      & 22          & * & 33 \\
10     & 30          & * & 40 \\
11     & 36          & * & 45 \\
12     & 42          &   & 42 \\
13     & 38          &   & 38 \\
14     & 36          &   & 36 \\
15     & 33          &   & 33 \\
16     & 31          &   & 31 \\
17     & 29          &   & 29 \\
\hline
{AveP} & 10          & * & 14 \\
\hline
\end{tabular}
\end{table}

The asterisks (*) in the table mark where the two methods have different results. 
The table shows that the influence-primed model identified the ACL fellows with a better
precision for seven values of $N$. When we limit $N$ to less than or equal to seventeen,
hip-index is never worse but often better than \mbox{$h$-index}. 
As $N$ grows, the differences between the two indexes become negligible. At $N=17$, both 
indexes identify 5 out 17~ACL Fellows.

The last row of Table~\ref{tab:fellow} shows the \emph{average precision} measure 
(AveP), which is commonly used to evaluate search engines~\cite{Buckley:2000}. The 
formula for AveP is 

\begin{equation*}
\textrm{AveP}(n_c) = \frac{\sum_{k=1}^{n_c}(P(k) \times \textrm{rel}(k))}{n_r} ,
\end{equation*}

\noindent where $n_c$ is the point at which we cut off the list of search
results for a search engine (in our case, $n_c = 17$, where we cut off the ranked
lists of authors), $n_r$ is the total number of relevant documents (in our case,
the total number of ACL Fellows, $n_r = 17$), $P(k)$ is the precision at each observation point 
(in our case, $k$ ranges from 1 to 17), and $\textrm{rel}(k)$ is an indicator function that equals 
1 if the \mbox{$k$-th} document is relevant to the given query and 0 otherwise (in our case,
$\textrm{rel}(k)$ is 1 if the \mbox{$k$-th} author is an ACL Fellow and 0 otherwise). 
From the table, we can see  that the AveP score of the influence-primed model is 14\% 
whereas that of the conventional model is 10\%.

This better average precision measure is encouraging evidence that weighting the citations 
by our measures of influence could improve the identification of the best scientists. 

\section*{Future work and limitations}

Further work is needed to validate these results over more extensive and different datasets. 
We rely on authors to annotate their own papers. However, we did not
assess the reliability of authors at identifying the key references.
For papers with several authors, we could ask more than one author to provide annotation. 
Thus we could quantify the inter-annotator agreement. We could also
ask the same authors, after a long delay, to annotate their own papers again.
Furthermore, we would find it interesting to compare the performance of a machine learning 
approach with human-level performance. For this purpose, we could recruit
independent annotators having specific degrees of expertise.

We assumed that the full text of the citing paper was available. Yet some
citation indexing databases have a more limited access to the content due to
copyright restrictions or technical limitations. It
should not be difficult to extend these databases so that they have the necessary 
information to identify influential references or to count the number of times a 
paper mentions another. For example, this information could be 
provided by the copyright owner without giving access to the full text.

Intentionally, we limited our feature set so that we did not have
to recover full text of the cited work (only the title). However, many other features are possible
if we access the full text of both the citer and cited paper: \citeA{BIES:BIES201100067} 
measure similarity by the overlap in the reference section. We could also add other features 
such as the prestige  of the cited venue or the prestige of the cited authors~\cite{zitt2008modifying}.
We also did not take into account the relationships between authors. Maybe
authors who are similar or related are more likely to influence each other.
In related work, \citeA{ajiferuke2010comparison} proposed to measure the number of 
citers (authors who cite) rather than the number of citations. 

Moreover, identifying the genuinely significant citations might be viewed as an 
adversarial problem. Indeed, some authors and editors attempt to game (i.e., manipulate 
or exploit) citations counts. In a survey, \citeA{wilhite2012coercive} found that 
20\% of all authors were coerced into citing some references by an editor, after their 
manuscript had undergone normal peer review. In fact, the majority of 
authors reported that they were willing to add superfluous citations if it is an 
implied requirement by an editor.  If we could determine that many non-influential 
references in some journals are citing some specific journals, this could indicate 
unethical behavior.

Our approach for identifying celebrated scientists is simple. State-of-the-art approaches
such as that of \citeA{rokach2011going} can achieve better precision and recall. 
We expect that they would benefit from an identification of the influential citations.

In some circumstances counting multiple occurrences of a citation might require 
coreference resolution~\cite{Soon:2001,Athar:2012:DIC:2391171.2391176}.  By convention, 
some authors and editors only ever cite a reference once but mention it several times 
either with a nominal, or pronominal reference.

Mazloumian~\cite{mazloumian2012predicting} found that a useful predictor of 
future performance was the annual  number of citations at the time of citation. 
Maybe the annual number of influential citations could be a superior predictor.

A finer view of the problem is to rank or rate the references by their degrees
of academic influence, which however could bring further complexity
that we are avoiding in the present work; e.g., comparing two less
influential or uninfluential references could be a harder task even
for human annotators, and such annotation may be difficult to interpret.

Nevertheless, a weighted citation measure based on 
the number of occurrences of a citation could significantly alter other
evaluation metrics that depend on simple citation counts, such as Impact 
Factor~\cite{Garfield:2006}, Eigenfactor, and Article Influence~\cite{bergstrom:2007}. 
Even though such a refinement would not address the \emph{statistical} charges 
leveled against citation-based evaluation metrics~\cite{Adler:2009}, it would  at 
least address to some degree the need to distinguish between citations that acknowledge 
an \emph{intellectual debt} and \emph{de-rigeur} citations.

We conjecture that some types of research papers that tend to be highly cited, such 
as review or methodological articles, are less likely to be perceived as influential. 
\citeA{BIES:BIES201100067} found that weighting citations by the in-paper frequency
reduced the importance of reviews. We should further investigate this issue to verify whether 
original contributions are significantly more likely to be perceived as influential.

\section*{Conclusions}

One of our main results is that counting the number of times a paper is cited 
($countsInPaper\_whole$) is one of the best predictors of how influential a reference is. 
This confirms an earlier result by \citeA{BIES:BIES201100067} who stated,
``Citation frequency of individual articles in other papers more fairly measures 
their scientific contribution than mere presence in reference lists.''
Alternatively, we can count the number of sections in which a paper is cited 
($countsInPaper\_secNum$). 

We believe that in assessing the influence of a research paper or researcher,
weighting the citations by these features (e.g., $countsInPaper\_whole$) would 
provide more robust results. It should also be used when tracking follow-up work 
or recommending research papers.

We have also shown that we could combine the in-paper
citation counts ($countsInPaper$) and the semantic relatedness between
a reference and the citing paper, to derive a superior classifier.
Though self-citations are only slightly correlated with academic influence, 
a classifier can derive some benefits when combining it with other features.

\section*{Acknowledgments}

We are grateful to the volunteers who identified key citations in their own work.
Daniel Lemire acknowledges support from the Natural Sciences and Engineering Research 
Council of Canada (NSERC) with grant number 26143. We thank M.~Couture, V.~Larivi\`ere 
and the anonymous reviewers for their helpful comments.

\theendnotes

\bibliographystyle{theapa}
\bibliography{affinity}

\end{document}